\documentclass [a4paper,fleqn, 12pt]{article}
\usepackage{graphicx}

\usepackage {amsmath} \usepackage{amssymb} \usepackage{cite}

\newcommand{\cn}

\begin{document}

\title{Extended two dimensional equation for the description of nonlinear waves in gas--liquid mixture}

\author{Nikolay A. Kudryashov, \and Dmitry I. Sinelshchikov \and Alexandr K. Volkov}

\date{Department of Applied Mathematics, National Research Nuclear University MEPHI, 31 Kashirskoe Shosse, 115409 Moscow, Russian Federation}

\maketitle

\begin{abstract}
We consider a system of equations for the description of nonlinear waves in a liquid with gas bubbles. Taking into account high order terms with respect to a small parameter, we derive a new nonlinear partial differential equation for the description of density perturbations of mixture in the two-dimensional case. We investigate integrability of this equation using the Painlev\'{e} approach. We show that travelling wave reduction of the equation is integrable under some conditions on parameters. Some exact solutions of the equation derived are constructed. We also perform numerical investigation of the nonlinear waves described by the derived equation.
\end{abstract}


 \section{Introduction}
 \label{introduction}
 A liquid with gas bubbles is a complex dissipative and dispersive nonlinear media. Nonlinear character of waves in such medium brings essential difficulties for investigation, although there are some interesting properties of wave processes in gas-liquid mixture and mathematical models for the description of such systems occurs widely in different sciences: chemistry, biology, physic and etc. (see \cite{Nigmatulin, Nakoryakov, Goldberg}).
 
 For the first time nonlinear evolution equations like Burgers, Korteweg--de Vries and Burgers--Korteweg--de Vries were obtained for the description of long weakly nonlinear waves in a gas--liquid mixture in works \cite{Winjngaarden1, Winjngaarden2, Nakoryakov2} for the one-dimensional case. The three--dimensional case was considered in work  \cite{three-dimensional}, but only first--order terms in an asymptotic series have been taken into account. On the other hand, considering high--order corrections in asymptotic expansions, we are able to obtain more complicated nonlinear equations. It allows us to describe wave processes more accurate then in \cite{three-dimensional}. Besides, we can discover some new physical effects. In work \cite{Extanded_models} models for non--linear waves in a gas--liquid mixture were generalized, taking into account higher order terms with respect to small parameters. Models of work \cite{Extanded_models} take into consideration an interphase heat transfer, surface tension and weak liquid compressibility, although only one--dimensional case is considered. Thus, it is interesting to study long weakly nonlinear waves in a liquid with a gas bubbles in two--dimensional case, taking into consideration both high order terms in the asymptotic expansions and physical properties mentioned above.
 
 Here we derive a new nonlinear partial differential equation for the description of long weakly nonlinear two--dimensional waves in a bubbly gas--liquid mixture. We consider waves propagating in a certain direction. We assume that perturbations in perpendicular directions are less essential but we take them into account. We take into account high order terms in the asymptotic expansions, interphase heat transfer, surface tension and weak liquid compressibility. We also investigate equation derived analytically and numerically. To the best of our knowledge, this equation has not been obtained and investigated before.
 
 In order to investigate integrability of the nonlinear equation we apply the Painlev\'{e} approach. It is shown that the equation does not have the Painlev\'{e} property in general case. However solitary wave solutions are constructed by means of the truncated expansion method. Using travelling wave variables it is shown that the equation passes the Painlev\'{e} test under some conditions on parameters. With the Hopf--Cole transformations the equation  is linearised and its solutions are obtained in different forms. Nonlinear waves described by the equation are also investigated numerically using the spectral method. It is shown, that this method have good accuracy and stability.
 
 The rest of this work is organized as follows. In section \ref{equation} we derive a nonlinear  partial differential equation for the description of waves in gas--liquid mixture, taking into consideration second order terms with respect to the small parameters.  In section \ref{general_case} we apply the Painlev? approach to investigate integrability of the equation. In section \ref{travelling_wave} the new nonlinear equation is investigated using travelling wave variables. It is shown that the equation is integrable under some conditions on parameters. In section \ref{numerical} we present the results of the numerical simulation of waves, described by the equation. In the last section we briefly discuss our results.

 
 \section{Extended equation for the description of waves in a liquid with gas bubbles in two--dimensional case.}
 \label{equation}
 In this section we obtain a two--dimensional nonlinear equation for the description of waves in a liquid with gas bubbles. We use the system of equations for the description of waves in bubbly liquid, presented in \cite{three-dimensional}. We suppose that the gas--liquid mixture is a homogeneous medium with an average pressure and temperature. We assume that the liquid is incompressible and gas bubbles are spherical. We do not consider destruction, formation, interaction and coalescence of bubbles. We suppose that total amount of gas in a bubble and the amount of gas bubbles in unit of mass of liquid are constants. Gas in bubble is an ideal and a pressure in bubble is described by the politropic law. Liquid viscosity is considered only on the interphase boundary. 
 Taking into account assumptions mentioned above, the following system of equations for the description of waves in liquid with gas bubbles is used (see \cite{three-dimensional} )
 \begin{equation}
 \begin{gathered}
 \label{start_system2}
 \frac {\partial{\tilde \rho}} {\partial{\tau}} + \nabla \tilde{\textbf u} + \nabla (\tilde \rho \tilde{\textbf u} ) = 0, \\
 (1+\tilde \rho)\left(\frac {\partial{\tilde {\textbf u} }} {\partial{\tau}} +  \tilde{\textbf u} \nabla \tilde{\textbf u} \right) + \frac 1 \alpha \nabla \tilde p = 0, \\
 p=\alpha \tilde \rho + \alpha_1 \tilde \rho^2 +\alpha_2 \tilde \rho^3 + \beta \tilde \rho_{\tau \tau}  - \left(\beta_1 + \beta_2\right)\tilde \rho \tilde \rho_{\tau \tau} - \left(\beta_1 + \frac 3 2 \beta_2\right) \tilde \rho_\tau^2 +  \varkappa \tilde \rho_\tau + \varkappa_1 \tilde \rho \tilde \rho_{\tau}.
 \end{gathered}
 \end{equation}
 Here $p, \, \tilde{\rho}, \, \tilde {\textbf u}$ are the non--dimensional pressure, density and velocity of the mixture correspondingly, $\xi,  \, \eta$ are Cartesian coordinates and $\tau$ is  the time; $\alpha, \, \alpha_1, \, \alpha_2, \, \beta, \, \beta_1, \, \beta_2, \, \varkappa, \, \varkappa_1 $ are non--dimensional parameters \cite{three-dimensional}. 
 
 For the derivation of an equation for the description of nonlinear waves, we use the reductive perturbation method (see e.g. \cite{Washimi, Gardner, Kako, Taniuti, Parkes}). Let us introduce 'slow' variables
 \begin{equation}
 \label{slow_var}
 x=\epsilon(\xi - \tau), \; y= \epsilon^{\frac32} \delta \eta, \; t=\epsilon^2\tau.
 \end{equation}
 We suppose that perturbations in $x$ direction are more essential then in $y$. We chose power of $\epsilon$ in 'slow' variables in order to obtain equations for the case of dissipation main influence. We search for a solution of system \eqref{start_system2} in the form of asymptotic series:
 
 \begin{equation}
 \begin{gathered}
 \label{sol_ser}
 \tilde u^{(1)} = \epsilon u^{(1)}_1 + \epsilon^2 u^{(1)}_2 + \cdots ,\qquad
 \tilde u^{(2)} = \epsilon u^{(2)}_1 + \epsilon^2 u^{(2)}_2 + \cdots,\\
 \tilde \rho = \epsilon \rho_1 + \epsilon^2 \rho_2 + \cdots,\qquad
 \tilde p = \epsilon p_1 + \epsilon^2 p_2  + \cdots .
 \end{gathered}
 \end{equation}
 Substituting \eqref{slow_var} and \eqref{sol_ser} into \eqref{start_system2} and collecting coefficients at $\epsilon^0$ we obtain
 \begin{equation}
 \label{2d1}
 u^{(1)}_1 = \rho_1, \qquad
 p_1 = \alpha \rho_1.
 \end{equation}
 
 Collecting coefficients at the same powers of $\epsilon$ and using \eqref{2d1} we have the following equations:
 \begin{equation}
 \begin{gathered}
 \label{5}
 \rho_{1t} - \rho_{2x} + u^{(1)}_{2x} + (\rho_1 u^{(1)}_1)_x + \epsilon \delta u^{(2)}_{1y}  = 0, \\
 u^{(1)}_{1t} - u^{(1)}_{2x} + u^{(1)}_1 u^{(1)}_{1x} + \rho_{2x} + \frac{2\alpha_1}\alpha \rho_1 \rho_{1x} - \frac \varkappa \alpha \rho_{1xx} - \rho_1 u^{(1)}_{1x}+ \\
 +\epsilon \left(\rho_1 u^{(1)}_{1t} + \rho_1u^{(1)}_1u^{(1)}_{1x} + \frac{\alpha_2}\alpha (\rho_1^3)_{x} + \frac \beta \alpha \rho_{1xxx} + \frac {\varkappa} \alpha \rho_{1tx} - \frac {\varkappa_1}\alpha (\rho_1 \rho_{1x})_x \right) = 0,
 \end{gathered}
 \end{equation}
 \begin{equation}
 \label{6}
 u^{(2)}_{1x} = \delta \rho_{1y} + \epsilon \left(u^{(2)}_{1t}+u^{(2)}_1u^{(1)}_{1x} \frac{\delta \alpha_1}{\alpha}\rho^2_{1y} -\frac{\delta\varkappa}{\alpha}\rho_{1xy} - \rho_1u^{(2)}_{1x}\right).
 \end{equation}
 Differentiating \eqref{5} and \eqref{6} with respect to $x$ and $y$ correspondingly and using obtained relations
 to avoid velocity, we get:
 \begin{equation}
 \begin{gathered}
 \label{rho_without_t}
 \left( \rho_{1t} + \left( 1 + \frac {\alpha_1 } \alpha \right) \rho_1 \rho_{1x} - \frac \varkappa {2\alpha} \rho_{1xx} \right)_x +\\
 +\epsilon \frac 1 2  \left( \frac{3\alpha_2 - \alpha_1}\alpha \rho_1^2\rho_{1x}
 + \left( \frac \beta \alpha +\frac{\varkappa^2}{2\alpha^2} \right) \rho_{1xxx}
 + \frac{\varkappa}{2\alpha} \rho_1\rho_{1xx}
 - \frac{\varkappa\left(2\alpha + \alpha_1 \right)}{\alpha^2} (\rho_1 \rho_{1x})_x \right)_x +\\
 + \epsilon \frac {\delta^2} 2 \rho_{1yy} = 0.
 \end{gathered}
 \end{equation}
 Now, using the near-identity transformations \cite{Kraenkel, Veksler}
 \begin{equation}
 \rho_1 = \rho + \epsilon\left( \lambda_1 \rho^2 + \lambda_2 \rho_x \partial_x^{-1}\rho\right),
 \end{equation}
 we obtain  equation 
 \begin{equation}
 \left( \rho_t + a_1\rho\rho_x + a_2 \rho_{xx} + a_3(\rho \rho_x)_x + a_4\rho^2\rho_x + a_5 \rho_{xxx} \right)_x + b\rho_{yy}=0,
 \end{equation}
 where
 \begin{equation}
 \begin{gathered}
 a_1=\left(1+\frac{\alpha_1}{\alpha}\right), \quad
 a_2 = - \frac{\varkappa}{2\alpha}, \quad
 a_3 = \epsilon \frac{\varkappa(1-2\alpha - \alpha_1) - 4\alpha(\lambda_1 + 2\lambda_2) }{4\alpha^2}, \\
 a_4 = \epsilon\left( \frac{3\alpha_2 - \alpha_1}{2\alpha} +  \left(1+\frac{\alpha_1}{\alpha}\right)\left(\lambda_1+\lambda_2\right) \right), \quad
 a_5 = \epsilon\frac{2\beta\alpha+\varkappa^2}{4\alpha^2}, \quad
 b = \epsilon \frac{\delta^2}{2}.
 \end{gathered}
 \end{equation}
 and 
 \begin{equation}
 \lambda_1 = - \lambda_2 + \frac{\varkappa \left(1-6\alpha-3\alpha_1 \right)}{\left(8\varkappa+4\right)\alpha}.
 \end{equation}
 Let us use shift and scaling transformations in the form:
 \begin{equation}
 \begin{gathered}
 \label{shift_and_scaling}
 x'=Ax+Bt, \quad
 t' = Ct, \quad
 \rho' = D\rho + E, \quad
 y' = Fy, 
 \end{gathered}
 \end{equation}
 where $A$ is an arbitrary parameter, $B,\; C,\; D,\; E,\; F$ are described by the following relations:
 \begin{equation}
 \begin{gathered}
 B = \frac{a_1^2A}{4a_4}, \quad
 C =  a_5 A^3, \quad
 D = -\frac{a_5A}{a_3}, \quad
 E = -\frac{a_1}{2a_4}, \quad
 F = \sqrt{\frac{a_5}{b}}A^2, \quad
 \mu = \frac{a_4a_5}{a_3^2},
 \end{gathered}
 \end{equation}
 constant $\lambda_2$ is found from the equation $a_1a_3 = 2a_2a_4$. Taking into account transformations \eqref{shift_and_scaling} we  
 can rewrite equation \eqref{rho_without_t} in the form
 \begin{equation}
 \left( \rho_t  + \mu \rho^2 \rho_x + \rho_{xxx} - (\rho \rho_x)_x \right)_x +  \rho_{yy}= 0.
 \label{double_dimension_equation}
 \end{equation}
 Below we study equation \eqref{double_dimension_equation}.
 
 \section{Painlev\'e test to equation \eqref{double_dimension_equation}}
 \label{general_case}
 
 To investigate integrability of \eqref{double_dimension_equation} let us apply the Weiss--Tabor--Carnevale (WTC) test \cite{Weiss, Weiss1}. We look for a solution of \eqref{double_dimension_equation} in the form of series
 \begin{equation}
 \label{solution_expansion2}
 \rho(x, y, t) = \Phi^p \sum_{j=0}^{\infty} u_j \Phi^j,
 \end{equation}
 where $\Phi=\Phi(x,y,t)$ is a new function and $u_j = u_j(x,y,t)$ are coefficients in expansion \eqref{solution_expansion2}. The leading terms of equation \eqref{double_dimension_equation} are $\mu \left( \rho^2 \rho_x \right)_x, \, \rho_{xxxx}$ and $-\left(\rho \rho_x \right)_{xx} $. Substituting $\rho = u_0 \Phi^{p}$ into leading terms and equating coefficients at the lowest order of $\Phi(x,y,t)$ we obtain
 
 \begin{equation}
 p = -1, \qquad u_0 =\frac{\left(-3\pm \sqrt{9-24\mu}\right)\Phi_x }{2\mu}.
 \end{equation}
 Then, substituting into leading terms the expression
 \begin{equation}
 \rho(x, y, t) = \frac{u_0(x,y,t)}{\Phi(x,y,t)} + u_{0j}\Phi(x,y,t)^{j-1},
 \end{equation}
 and equating the coefficient at $u_j$ to zero, the following Fuchs indexes are found for different branches of solution
 \begin{equation}
 \begin{gathered}
 j_1^{(1,2)}=-1,	\quad j_2^{(1,2)}=3, \quad j_3^{(1,2)}=4, \quad
 j_4^{(1,2)} = {\frac {\pm \sqrt {9-24\,\mu}+8\,\mu-3}{2 \mu}}.
 \label{Fuchs_indexes}
 \end{gathered}
 \end{equation}
 Here upper index is a number of a branch of the solution and lower index is a number of an index for definite branch.
 Let us recall that an equation pass the Painlev\'e test only if Fuchs indexes are integer. This holds only for the limited number of values of $\mu$. In case of $\mu=\frac38$ we have the flowing Fuchs indexes on each branch of the solution:
 \begin{equation}
 \begin{gathered}
 j_1^{(1)}=-1,\qquad j_2^{(1)}=0,\qquad j_3^{(1)}=3,\qquad j_4^{(1)}=4,\\
 j_1^{(2)}=-1,\qquad j_2^{(2)}=0,\qquad j_3^{(2)}=3,\qquad j_4^{(2)}=4.
 \end{gathered}
 \end{equation}
 We see that equation \eqref{double_dimension_equation} does not pass Painlev\'e test in case of $\mu=\frac38$, because if the Fuchs index equal to zero, $u_0$ must be an arbitrary function, but it is determined, thus we can not take three arbitrary constants in expression \eqref{solution_expansion2}. 
 
 In case of $\mu=-3$ we have two sets of Fuchs indexes:
 \begin{equation}
 \begin{gathered}
 j_1^{(1)}=-1,\qquad j_2^{(1)}=3,\qquad j_3^{(1)}=3,\qquad j_4^{(1)}=4;\\
 j_1^{(2)}=-1,\qquad j_2^{(2)}=3,\qquad j_3^{(2)}=4,\qquad j_4^{(2)}=6.
 \end{gathered}
 \end{equation}
 Thus equation \eqref{double_dimension_equation} does not pass Painlev\'e test when $\mu=-3$ because $j_4^{(1)}=j_2^{(1)}=3$ and expression \eqref{solution_expansion2} has to have logarithmic terms. 
 
 Let us consider case $\mu = \frac13$. We have Fuchs indexes
 \begin{equation}
 \begin{gathered}
 j_1^{(1)}=-1,\qquad j_2^{(1)}=1,\qquad j_3^{(1)}=3,\qquad j_4^{(1)}=4,\\
 j_1^{(2)}=-1,\qquad j_2^{(2)}=-2,\qquad j_3^{(2)}=3,\qquad j_4^{(2)}=4.
 \end{gathered}
 \end{equation}
 According to the idea of Painlev\'{e} test we recall that \eqref{double_dimension_equation} have the Painlev\'{e} property if  functions $\Phi, \,u_1, \, u_3, \, u_4$ are arbitrary in \eqref{solution_expansion2} for one branch of solution and for the other branch we need to use special modification of Painlev\'{e} test. Substituting \eqref{solution_expansion2}  into \eqref{double_dimension_equation} and  consistently equating terms at various powers of $\Phi$ to zero, we find
 \begin{equation}
 \begin{gathered}
 u_0 = -3\Phi_x, \qquad
 u_2 = \frac{u_1^2}{3\Phi_x} - \frac{u_{1x}}{\Phi_x} - \frac{\Phi_{xx}u_1}{\Phi_x^2} + \frac{\Phi_y^2}{\Phi_x^3} + \frac{\Phi_{xxx}}{\Phi_x^2} + \frac{\Phi_t}{\Phi_x^2},
 \end{gathered}
 \end{equation}
 and $u_1$ is an arbitrary function. Coefficient at $\Phi^{-2}$ has to be zero as $u_3$ has to be an arbitrary function. However we obtain that $u_3$ can be arbitrary only if the following equation is satisfied:
 \begin{equation}
 \label{penleve_coeff}
 3\frac{(\Phi_y)^2 \Phi_{xx}}{\Phi_x} + 3\Phi_{yy} \Phi_x - 6\Phi_y \Phi_{xy} = 0.
 \end{equation}
 It is obvious from \eqref{penleve_coeff} that $u_3$ can not be taken as an arbitrary function. Thus equation \eqref{double_dimension_equation} does not have Painlev\'{e} property
 and seems not to be integrable. However one is able to find a solution for equation \eqref{double_dimension_equation} applying truncated expansion method \cite{Kudryashov90, Kudryashov91}. We search for a solution of equation \eqref{double_dimension_equation} in the form
 \begin{equation}
 \label{trunc_sol}
 \rho(x,y,t) = \frac{A_0(x,y,t)}{\Phi(x,y,t)}+A_1(x,y,t).
 \end{equation}
 Substitute expression \eqref{trunc_sol} into equation \eqref{double_dimension_equation} and equating coefficients at different powers of $\Phi$ to zero we find coefficients $A_0, A_1$. As general case leads to cumbersome formulas we fix parameter $\mu=\frac38$, $\mu=-3$ or $\mu=\frac13$. These cases correspond to integer values of Fuchs indexes. We also search for the solution in the form of solitary wave so we set $\Phi = e^{k_x x + k_y y - \omega t + \phi_0}$. After all operations in case of $\mu=\frac38$ we get the dispersion relation
 \begin{equation}
 \omega = \frac{1}{2 k_1}\left(k_x ^4 + 2 k_y^2 \right),
 \end{equation}
 and functions
 \begin{equation}
 \begin{gathered}
 A_0 = -4k_x e^{k_x x + k_y y - \omega t + \phi_0}, \;
 A_1 = 2k_x.
 \end{gathered}
 \end{equation}
 
 Thus, the solution of \eqref{double_dimension_equation} in the form of solitary wave have the form
 \begin{equation}
 \label{soliton_sol_mu38}
 \rho=-2k_x\frac{ \left(e^{k_x x + k_y y - \omega t + \phi_0} -1 \right)}{1+e^{k_x x + k_y y - \omega t + \phi_0}}.
 \end{equation}
 
\begin{figure}[!ht]
 	\centering
 	\includegraphics[width=0.60\textwidth]{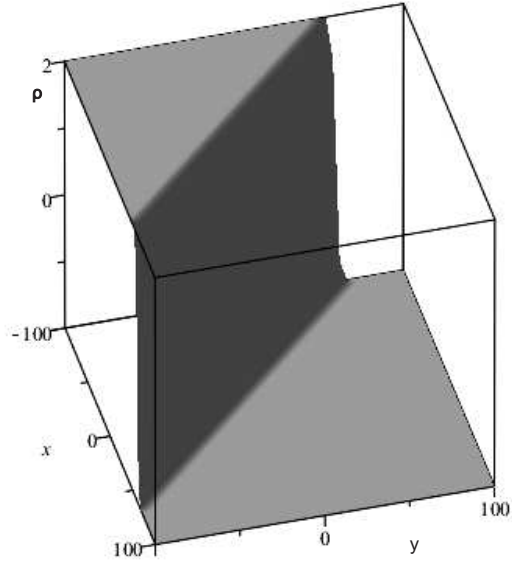}	
 	\caption{one--soliton solution \eqref{soliton_sol_mu38} at time moment $t=0, $ as $k_x = 1, \; k_y = 1,\; \phi_0= 40$}
 	\label{cut_sol_m38}
 \end{figure}

 On figure \ref{cut_sol_m38} the exact solution for the equation \eqref{double_dimension_equation} in the form of solitary wave \eqref{soliton_sol_mu38}  is illustrated.
 
 Analogously to case of $\mu =  \frac38$, setting $\mu=-3$ dispersion relation and parameters take the form
 \begin{equation}
 \begin{gathered}
 \label{param_mu-3}
 \omega = -\frac{1}{4 k_x}\left(k_x ^4 - 4 k_y^2 \right), \quad
 A_0 = -k_x e^{k_x x + k_y y - \omega t + \phi_0}, \quad
 A_1 = \frac{k_x}2,
 \end{gathered}
 \end{equation}
 Substituting \eqref{param_mu-3} into \eqref{trunc_sol} analytical solution of the equation \eqref{double_dimension_equation}  is found in the form
 \begin{equation}
 \label{soliton_sol_mu-3}
 \rho=-\frac{k_x}{2} \frac{ \left( e^{k_x x + k_y y - \omega t + \phi_0} -1\right)}{1+e^{k_x x + k_y y - \omega t + \phi_0}}.
 \end{equation}
 And in case of $\mu = \frac13$ one can obtain relations
 \begin{equation}
 \begin{gathered}
 \label{param_mu13}
 \omega = \frac{k_x ^4 + k_y^2}{k_x}, \qquad
 A_0 = -6k_x e^{k_x x + k_y y - \omega t + \phi_0}, \qquad
 A_1 = 3k_x,
 \end{gathered}
 \end{equation}
 and the solution of equation \eqref{double_dimension_equation}
 \begin{equation}
 \label{soliton_sol_mu13}
 \rho=-3k_x \frac{ \left( e^{k_x x + k_y y - \omega t + \phi_0} -1\right)}{1+e^{k_x x + k_y y - \omega t + \phi_0}}.
 \end{equation}
 It is obvious that solutions \eqref{soliton_sol_mu-3} and \eqref{soliton_sol_mu13} has the same type as \eqref{soliton_sol_mu38}.
 
 \section{Travelling wave solutions of equation \eqref{double_dimension_equation}.}
 \label{travelling_wave}
 
 Let us investigate equation \eqref{double_dimension_equation} using the travelling wave variables. Assuming that $\rho=\rho(z),\; z=x+y-C_0t$
 and integrating equation \eqref{double_dimension_equation} with respect to $z$, we get
 \begin{equation}
 \label{trav_wave_eq}
 (1-C_0)\rho_{z} +  \mu \rho^2 \rho_{z} +  \rho_{zzz} -  (\rho \rho_{z})_z  + C_1= 0,
 \end{equation} 
 where $C_1$ is a constant of integrating. Substituting $\rho(z) = a_0z^p + a_jz^{p+j}$ into the leading terms of equation \eqref{trav_wave_eq}, we find constants $p=1$, $a_0 = -\frac {3 \pm \sqrt {9-24\,\mu}}{2 \mu}$ and three Fuchs indexes which are the same as \eqref{Fuchs_indexes} except index $j=4$. Therefore equation \eqref{trav_wave_eq} may pass the Painlev\'e test only if $\mu=\frac13$. In this case the Fuchs indexes are 
 \begin{equation}
 \begin{gathered}
 j^{(1)}_1=-1,\qquad j^{(1)}_2=1,\qquad j^{(1)}_3=3; \\
 j^{(2)}_1=-1,\qquad j^{(2)}_2=-2,\qquad j^{(2)}_3=3.
 \end{gathered}
 \end{equation}
 Here upper index stands for the number of branch of solution of \eqref{trav_wave_eq} and lower index is the number of index on the certain branch.
 On the first branch, according to Painlev\'e approach solution of equation \eqref{trav_wave_eq} is being searched in the form of expansion
 \begin{equation}
 \label{solution_expansion1d}
 \rho (z) = z^p \sum_{j=0}^{\infty} a_j z^j.
 \end{equation}
 Substituting \eqref{solution_expansion1d} into \eqref{trav_wave_eq} and equating coefficients at various powers of $z$ to zero, we obtain that $a_1,\; a_3$ are arbitrary constants and
 \begin{equation}
 a_2=\frac{a_1^2}{3} - C_0+1.
 \end{equation}
 The other constants can be defined too. On the second branch of solution there is the Fuchs index below zero.
 One can confirm that equation \eqref{double_dimension_equation}
 pass the Painlev\'e test in travelling wave variable using Conte--Fordy--Pickering algorithm (see, \cite{Conte1, Conte2}).
 
\begin{figure}[h]
 	\centering
 	\includegraphics[width=0.4\textwidth]{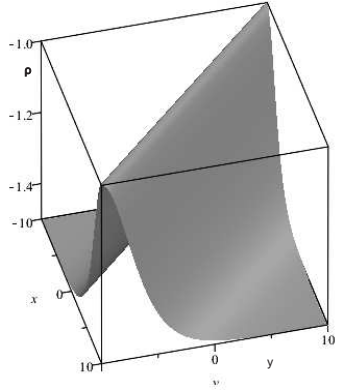} \hfil
 	\caption{solution of equation \eqref{trav_wave_eq} in case of $c_0=1, c_1=1, c_2=1, C_0=1.25$}
 	\label{Hopf-Cole2d_Gauss}
 \end{figure}

 Let us apply Hopf--Cole transformations $\rho(z)=-3 \frac{\psi'(z)}{\psi(z)}$ for linearization of equation \eqref{trav_wave_eq} in case of $\mu=\frac13$. Using this transformations we obtain linear equation for $\psi(z)$
 \begin{equation}
 \label{equation_for_psi}
 (1-C_0)\psi' + \psi''' = 0.
 \end{equation}
 Taking into account the general solution of equation \eqref{equation_for_psi} we obtain general solution of equation \eqref{trav_wave_eq}
 \begin{equation}
 \label{Hopf-Cole_solution}
 \rho(z) = \frac{ c_1 \sqrt{C_0-1} e^{\sqrt{c-1}z} + c_2 \sqrt{C_0-1}e^{-\sqrt{C_0-1}z  } }
 { c_0 + c_1 e^{\sqrt{C_0-1}z} +c_2 e^{-\sqrt{C_0-1}z} }.
 \end{equation}
 Under some conditions on parameters, for example $c_0=1, c_1=1, c_2=1, C_0=1.25$ we are able to obtain solution in the form of solitary wave. The result is demonstrated on figure \ref{Hopf-Cole2d_Gauss}. 
 
\begin{figure}[!h]
 	\centering
 	\includegraphics[width=0.4\textwidth]{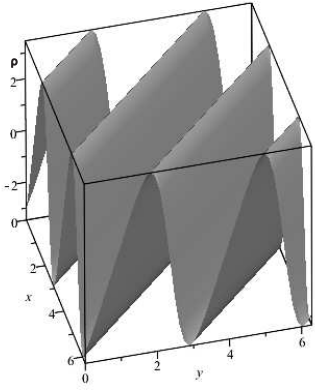}		\hfil
 	\caption{solution of equation \eqref{trav_wave_eq} in case of $t=0,\; C_0 = -3,\; c_0=1,\; c_1=1,\; c_2=1$}
 	\label{Hopf-Cole2d_periodic}
 \end{figure}

 When $c-1$ is below zero, we are able to obtain the periodical solution. In case of $t=0,\; C_0 = -3,\; c_0=1,\; c_1=1,\; c_2=1$ solution have form
 \begin{equation}
 \rho(x, y, t) = -\frac{6 \cos(6t+2x+2y)}{(2+\sin(6t+2x+2y))},
 \label{double_dim_for_num}
 \end{equation} 
 and is given on figure \ref{Hopf-Cole2d_periodic}.
 It is important to obtain solution of equation \eqref{double_dimension_equation} in the form of periodical wave \eqref{double_dim_for_num} because it can be used for the testing of program for numerical solution of equation \eqref{double_dimension_equation} with the spectral method and periodical boundary value conditions.
 
 Solution \eqref{Hopf-Cole_solution} is obtained under strong conditions on parameters. Let us apply method of the logistic function \cite{Kudryashov2012, Kudryashov2015} to get solution of equation \eqref{trav_wave_eq} in case of arbitrary value of parameter $\mu$. We search for the solution of equation \eqref{trav_wave_eq} in the following form:
 \begin{equation}
 \label{solution_form_logistic_function}
 \rho(z) = B_0 + B_1\Theta\left(z\right),\qquad \Theta\left(z\right) = \frac{1}{1-e^{-kz}},
 \end{equation}
 where $B_0,\; B_1$ are constants and $\Theta\left(z\right)$ is the so--called logistic function. Following \cite{Kudryashov2012}, we expend \eqref{solution_form_logistic_function} into Laurent series, substitute obtained expansion into equation \eqref{trav_wave_eq} and equate coefficients at different powers of $z$ to zero. We find that there are two solutions of equation \eqref{trav_wave_eq} in the form \eqref{solution_form_logistic_function} whith the following constants:
 \begin{equation}
 \begin{gathered}
 B^{(1,2)}_1 = \frac{(-3 \pm \sqrt{9-25\mu})k}{2\mu},\qquad
 B^{(1,2)}_0 =  \frac{(3 \mp \sqrt{9-25\mu})k}{4\mu},\\
 C^{(1,2)}_0 = \frac{\mp\sqrt{9-24\mu}k^2 - 4k^2\mu + 3k^2 + 8\mu}{8\mu}.
 \end{gathered}
 \end{equation}
 It is worth noting that since $B_1^{(1)} + B_1^{(2)} \neq 0$, there are no elliptic solutions of equation \eqref{trav_wave_eq}. We can also assume that $kz \rightarrow kz + i\pi$ and obtain solution in the form \eqref{solution_form_logistic_function} without poles on the real line.

\section{Numerical periodical solutions of the equation \eqref{double_dimension_equation}.}
\label{numerical}
In this section we study numerically nonlinear waves governed by equation \eqref{double_dimension_equation}. To this end we consider boundary value problem with periodical boundary conditions. Let us rewrite eq. \eqref{double_dimension_equation} in the form
\begin{equation}
\begin{gathered}
\label{numerical_problem}
\rho_{t} = L(\rho) + N(\rho), \\
\end{gathered}
\end{equation}
where $L(\rho)=-\rho_{xxx} - \partial^{-1}_x\rho_{yy}$ and $N(\rho)=\left(\rho\rho_x\right)_{x} - \mu\left(\rho^2\rho_x\right)$ are linear and nonlinear operator of the equation \eqref{double_dimension_equation} correspondingly. Antiderivative $\partial^{-1}_x$ is an integral with respect to $x$ . To solve boundary value problem which includes equation \eqref{numerical_problem}, start conditions and periodical boundary conditions, we use the integrating factor with the fourth--order Runge--Kutta approximation method (IFRK4), presented in works \cite{Cox, Kassam}. Using the Fourier transformation, we discretize the spatial part of equation \eqref{double_dimension_equation} and get the system of ordinary differential equations
\begin{equation}
\widehat{\rho}_t = \widehat{L(\rho)}+\widehat{N(\rho)},\qquad
\rho(x, 0) = \rho_0(x),
\end{equation}
where $\widehat{\rho},\; \widehat{L} = ik_x^3 -\frac{ik_y^2}{k_x}$ and $\widehat{N[\rho]} = -\frac{ik_x\mu}{3}\widehat{\rho^3} - \frac12k_x^2\widehat{\rho^2} $ are the Fourier forms of $\rho,\; L,\; N$ correspondingly, $k_x,\; k_y$ are Fourier multipliers. The basic idea of integrating factor (IF) method is to use transformations that allows us to solve the linear part of our problem exactly and then to solve the nonlinear part of our problem numerically. These transformations are the following:
\begin{equation}
\widehat{\rho}=e^{-\widehat{L}t}\rho,
\end{equation}
where $e^{-\widehat{L}t}$ is the IF. As a result, we obtain ordinary differential equation in the following form
\begin{equation}
\label{numerical_ode}
\widehat{\rho_t}=e^{-\widehat{L}t}\widehat{N}\left(\widehat{\rho}e^{-\widehat{L}t}\right).
\end{equation}
Equation \eqref{numerical_ode} with initial conditions can be solved by the fourth--ordered Runge--Kutta approximation method. It is obvious that equation \eqref{numerical_ode} has to be regularized for $k_x=0$  in order to give numerical sense to $\frac1k_x$. Following \cite{Klein}, we add to $k_x$ in the denominator a small imaginary part of appropriate sign $i\lambda_0$. For $\lambda_0$ we use the smallest floating point number that MATLAB can represent $2.2\times10^{-16}$. In this case the linear operator has form $\widehat{L} = ik_x^3 -\frac{ik_y^2}{k_x+i\lambda}$.

\begin{figure}[!ht]
	\centering
	\includegraphics[width=1\textwidth]{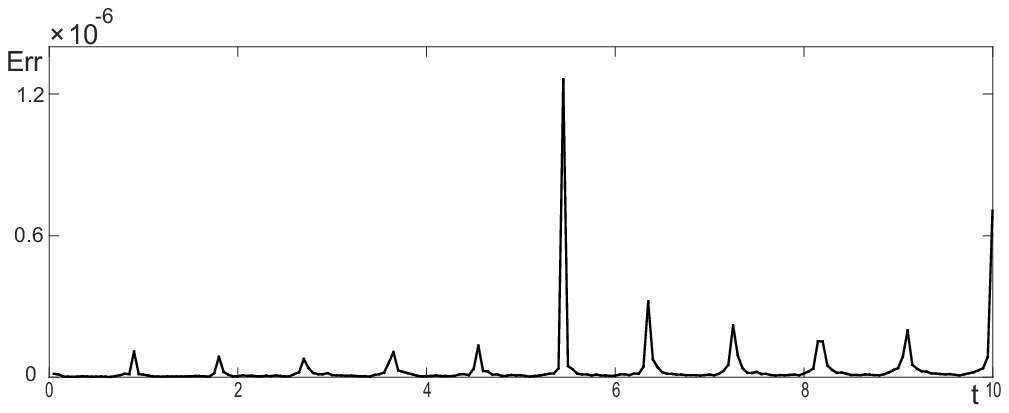}	
	\caption{evolution of the error $Err$ with the time $t$}
	\label{double_dim_err}
\end{figure}

To check our numerical strategy we used the exact solution of the equation \eqref{double_dimension_equation} in the periodical form \eqref{double_dim_for_num}. On each tome layer we define error for numerical simulation as
\begin{equation}
Err = \max_{i,j} \left|{\rho^{i,j}_{num} - \rho^{i,j}_{exact}}\right|,
\end{equation}
where $i$ and $j$ are discrate coordinates of points in $x$ and $y$ axis correspondingly, $\rho^{i,j}_{num}$ and $\rho^{i,j}_{exact}$ are values of a numerical and exact solutions in the grid point with the coordinate $(i, j)$ correspondingly. Evolution of the error with the time $t$ is given on the figure \ref{double_dim_err}. Taking into account results of numerical simulation of the exact solution we can regard the algorithm effective.

Let us use numerical simulation for the investigation of stability of equation \eqref{double_dimension_equation} to perturbations of the parameter $\mu$. In case fo $\mu=\frac13$, traveling wave reduction of equation \eqref{double_dimension_equation} is integrable and we use its exact solution \eqref{double_dim_for_num} as initial conditions for numerical simulation of equation \eqref{double_dimension_equation} in case of $\mu = \frac13 + \delta$. The result is presented on figure \ref{mu=1,3} for $\mu = 0.43$. We see that wave hold its shape but acquire different speed.

\begin{figure}[th!]
	\centering
	\includegraphics[width=1\textwidth]{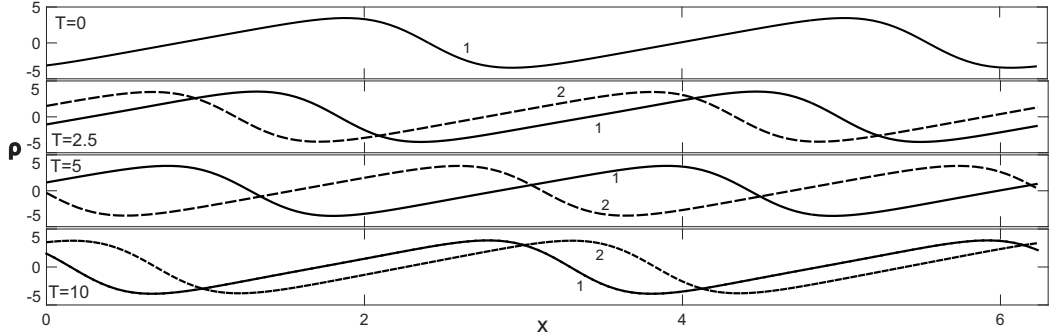}
	\caption{numerical solution of equation \eqref{double_dimension_equation} with periodical boundary conditions at $\mu=0.43$ (curve 1) and $\mu=\frac13$ (curve 2)}
	\label{mu=1,3}
\end{figure}

\begin{figure}[h]
	\centering
	\includegraphics[width=1\linewidth]{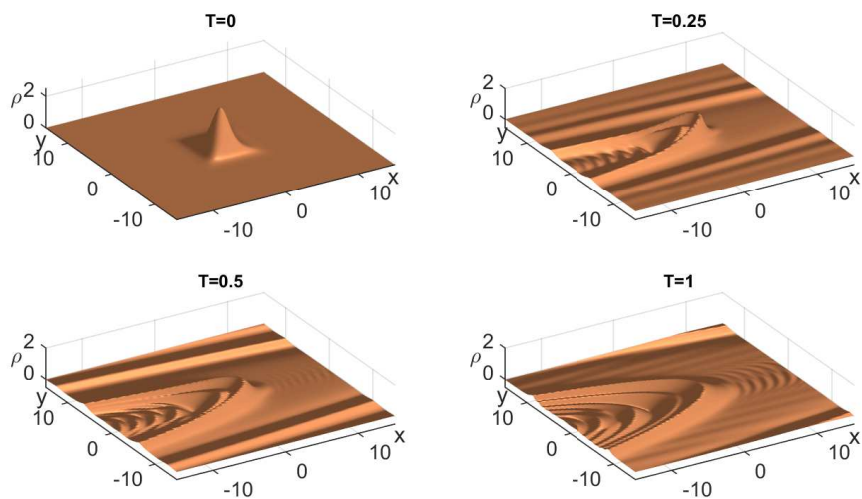}
	\caption[solitary wave simulation]{solitary wave simulation at $\mu=10$}
	\label{fig:double_dim_gauss}
\end{figure}

We also use solitary wave $\frac{5}{\cosh^2(x)+\cosh^2(y)}$ as an initial condition for numerical calculations. Result of calculations is presented on figure \ref{fig:double_dim_gauss}. Small perturbations in $y$ direction appears and amplitude of the solitary wave decreases during time evolution.

\section{Conclusion}
We have obtained new nonlinear equation \eqref{double_dimension_equation} for the description of waves in a liquid with gas bubbles in two--dimensional case. Using the Painlev\'e approach, we have shown that this equation is not integrable. We have constructed some analytic solutions of equation \eqref{double_dimension_equation} with the help of the truncated expansion method. We have shown that traveling wave reduction of eq. \eqref{double_dimension_equation} can pass the Painlev\'e test under some conditions on parameters.  We have shown that in this case equation \eqref{double_dimension_equation} can be linearized with the Hopf--Cole transformation. It has been shown that the equation for the description of waves in a gas--liquid mixture admits kink--type and periodical solutions. We have numerically investigated the evolution and stability of nonlinear waves discribed by equation \eqref{double_dimension_equation}.

\section{Acknowledgments}
This research was partially supported by grant for Scientific Schools 2296.2014.1., by RFBR grant 14--01--00498 and by
grant for the state support of young Russian scientists 3694.2014.1.

\end{document}